\documentclass[useAMS,usenatbib]{mnras}

\usepackage{aas_macros}
\usepackage{amsmath}
\usepackage{amssymb}
\usepackage{float}
\usepackage{graphicx}
\usepackage{lscape}
\usepackage{url}
\usepackage{natbib}
\usepackage{placeins}
\usepackage{color}
\usepackage{wasysym}
\usepackage{hyperref}
\usepackage{scrextend}
\DeclareMathOperator\erf{erf}
\begin{document}

\title{Generation of a Circumstellar Gas Disk by Hot Jupiter WASP-12b}

\author[A. Debrecht, J. Carroll-Nellenback, A. Frank, L. Fossati, E.G. Blackman \& I. Dobbs-Dixon]
{
Alex Debrecht$^{1}$, Jonathan Carroll-Nellenback$^{2}$, Adam Frank$^{1}$, Luca Fossati$^{3}$,\newauthor Eric G. Blackman$^{1}$ and Ian Dobbs-Dixon$^{4}$\\
$^{1}$Department of Physics and Astronomy, University of Rochester, Rochester NY 14627\\
$^{2}$Center for Integrated Research Computing, University of Rochester, Rochester NY 14627\\
$^{3}$Space Research Institute, Austrian Academy of Sciences, Schmiedlstrasse 6, A-8042 Graz, Austria\\
$^{4}$New York University Abu Dhabi, Abu Dhabi, United Arab Emirates
}

\date{}

\pagerange{\pageref{firstpage}--\pageref{lastpage}}
\maketitle
\label{firstpage}

\begin{abstract}

Observations of transiting extra-solar planets provide rich sources of data for probing the in-system environment. In the WASP-12 system, a broad depression in the usually-bright MgII h\&k lines has been observed, in addition to atmospheric escape from the extremely hot Jupiter WASP-12b. It has been hypothesized that a translucent circumstellar cloud is formed by the outflow from the planet, causing the observed signatures. We perform 3D hydrodynamic simulations of the full system environment of WASP-12, injecting a planetary wind and stellar wind from their respective surfaces. We find that a torus of density high enough to account for the lack of MgII\,h\&k line core emission in WASP-12 can be formed in approximately 13 years. We also perform synthetic observations of the Lyman-alpha spectrum at different points in the planet's orbit, which demonstrate that significant absorption occurs at all points in the orbit, not just during transits, as suggested by the observations.

\end{abstract}


\section{Introduction}
The discovery of extra-solar planets (exoplanets) has dramatically changed our understanding of planetary systems. Among these newly discovered systems, the transiting exoplanets are of particular importance, as they give us the unique opportunity to measure both planetary mass and radius (hence estimate their bulk density). These systems also allow characterization of planetary atmospheres through transmission/emission spectroscopy. Among the systems known to host a transiting close-in giant planet (a "hot Jupiter"), WASP-12 plays a special role: it is one of the first discovered inflated transiting hot Jupiters orbiting at an exceptionally close distance from the surface of a late F-type star \citep{hebb2009}. Because of the short orbital separation (only 1 stellar diameter from the stellar surface) and the effective temperature of the host star, WASP-12b's equilibrium temperature is of the order of 2500\,K. This is one of the highest among all known hot Jupiters, which has caused rising interest in this system over the years.

Together with HD209458b and HD189733b, WASP-12b is one of three hot Jupiters for which atmospheric escape has been directly observed through ultraviolet (UV) transit observations  \citep{vidalmadjar03,vidalmadjar04,vidalmadjar13,linsky10,lecavelier10,lecavelier12,fossati10a,haswell12}\footnote{Note that there is a tentative detection of an escaping atmosphere for the non-transiting hot Jupiter 55\,Cnc\,b \citep{ehrenreich2012}.}. Hubble Space Telescope (HST) near-UV transit observations of WASP-12b, carried out with the Cosmic Origins Spectrograph (COS), revealed the presence of transit depths up to three times greater than in the optical, clearly indicating that the planet's upper atmosphere overfills the Roche lobe and that the planet is therefore losing mass rapidly \citep{fossati10a,haswell12}. The data also allowed the detection of singly ionized Mg and Fe in the planet's upper atmosphere \citep{fossati10a,haswell12}, most likely dragged along by frequent collisions with H during the hydrodynamic escape \citep[e.g.][]{koskinen2013a,koskinen2013b,koskinen2014}, and of an early ingress compared to the optical ephemeris \citep{fossati10a,haswell12,nichols2015}. The early ingress phenomenon is currently believed to be caused by absorption from a bow-shock region ahead of the planet and supported by either the intrinsic planetary magnetic field or by the expanding planetary atmosphere \citep{vidotto2010,vidotto2015,bisikalo13}. A further possible explanation of the early-ingress phenomenon that has been put forward by \citet{lai2010} is the presence of a high-density (hence optically thick) area ahead of the planet where the stripped planetary material meets the torus.

Of significant interest, however, is the fact that the HST data also revealed WASP-12's stellar near-UV spectrum to present a broad depression in place of the normally bright emission cores in the MgII\,h\&k resonance lines \citep{haswell12}. This anomaly, which is always present regardless of the planet's orbital phase \citep[also in the core of the CaII\,H\&K resonance lines;][]{fossati13}, was completely unexpected given the spectral type and age of the star \citep{fossati2010b,haswell12}. \citet{fossati13} analysed the possible origin of this anomaly and first ruled out the hypothesis that WASP-12 presents an intrinsically and anomalously low stellar activity. This conclusion has recently been independently confirmed by the amount of jitter present in the stellar radial velocity measurements \citep{bonomo2017}. By directly measuring the ISM absorption from an early-type star lying close to the WASP-12 line of sight, \citet{fossati13} also excluded ISM absorption as the cause of the anomalous lack of line core emission.

Extrinsic absorption by material local to the WASP-12 system is therefore the most likely cause of the line core anomalies: gas escaping from the heavily irradiated planet could form a stable and translucent circumstellar cloud, as originally suggested by \citet{haswell12}. So far, no alternative explanation has been put forward; rather, the hypothesis of the presence of a translucent circumstellar cloud has been strengthened by theoretical and observational results \citep[e.g.,][]{lanza2014,fossati2015a}.

Recent global (star + planet) simulation studies of planet winds driven by stellar fluxes have demonstrated that escaping planetary material can collect around the star \citep{matsakos15,carroll16}.  In particular, in \citet{carroll16} it was shown that when radiation pressure is ineffective at driving material to larger radii \citep{Schneiter2007,Schneiter2017}, a torus of planetary wind material is created as strong Coriolis forces redirect wind parcels into up-orbit and down-orbit columns. These simulations indicate that it may be possible to build up dense enough torii around hot gas giants to produce the absorption seen in WASP-12b. 

To explore the possibility and origin of an absorbing circumstellar torus, we present here 3-dimensional (3D) hydrodynamic simulations of the whole WASP-12 system. We remark here that the only previous 3D hydrodynamic simulations of this system, presented by \citet{bisikalo13}, focused on the planet and the region immediately surrounding it. Our larger modelling domain allows us to study the fate of the material lost by the planet, and in particular if and how the escaped planetary material accumulates around the star. 

This paper is organized as follows. In Section \ref{sec:meth}, we discuss the model system and simulation parameters. In Section \ref{sec:result}, we present the results of the simulation. Our work is summarized and future avenues proposed in Section \ref{sec:discuss}.

\section{Methods and Model} \label{sec:meth}

Our simulations were conducted using AstroBEAR\footnote{https://astrobear.pas.rochester.edu/} \citep{cunningham09,carroll13}, a massively parallelized adaptive mesh refinement (AMR) code that includes a variety of multiphysics solvers, such as self-gravity, heat conduction, magnetic resistivity, radiative transport, and ionization dynamics. The equations solved for these simulations are those of fluid dynamics in a rotating reference frame, with gravitational effects of both the planet and star included:
\begin{equation}
    \frac{\partial \rho}{\partial t} + \boldsymbol{\nabla} \cdot \rho \boldsymbol{v} = 0 
    \label{eq:Eu1}
\end{equation}
\begin{equation}
    \frac{\partial \rho \boldsymbol{v}}{\partial t} + \boldsymbol{\nabla} \cdot \left ( \rho \boldsymbol{v} \otimes \boldsymbol{v} \right )= - \nabla p - \rho \nabla \phi + \boldsymbol{f_R}
    \label{eq:Eu2}
\end{equation}
\begin{equation}
    \frac{\partial E}{\partial t} + \boldsymbol{\nabla} \cdot ((E + p) \boldsymbol{v}) = 0
    \label{eq:Eu3}
\end{equation}
where $\rho$ is the mass density, $\boldsymbol{v}$ is the fluid velocity, $p$ is the thermal pressure, $\phi$ is the gravitational potential, $\boldsymbol{f_R}$ combines the the Coriolis and centrifugal forces, so that $\boldsymbol{f_R} = \rho \left ( - 2 \boldsymbol{\Omega} \times \boldsymbol{v} - \boldsymbol{\Omega} \times \left ( \boldsymbol{\Omega} \times \boldsymbol{r} \right ) \right )$ (where $\boldsymbol{\Omega}$ is the orbital velocity), and $E = \frac{1}{\gamma - 1} p + \frac{1}{2}\rho v^2$ is the combined internal and kinetic energies. To maintain the temperature of the stellar and planetary material, we chose a nearly isothermal value of $1.0001$ for $\gamma$.

\subsection{Model System Parameters}

The input parameters of the simulation were chosen on the basis of the WASP-12 system parameters, namely a planet with a radius of $1.736 R_{\jupiter}$ and mass of $1.404 M_{\jupiter}$, orbiting a late F-type star with a radius of $1.6 R_{\astrosun}$ and mass of $1.35 M_{\astrosun}$ at a separation of $a = 0.023 \mbox{AU}$ \footnote{http://exoplanet.eu}. Table \ref{tab:parameters} lists the parameters used in the simulations.

\setcounter{table}{0}

\begin{table}
\begin{minipage}{80mm}
\renewcommand{\thempfootnote}{\fnsymbol{footnote}}

\caption{Parameters used in the simulations}
\label{tab:parameters}
\begin{tabular}{l|c|c|}\hline

Planet Mass &$M_p$ & $1.404 M_{\jupiter}$ \footnote{\label{exoplanet} Exoplanet Encyclopedia (http://exoplanet.eu)} \\
Planet Radius &$R_p$ & $1.736 R_{\jupiter}$ \footref{exoplanet} \\
Planet Temperature &$T_p$ & $10^4 K$ \\
Planet Escape Parameter &$\lambda_p$ & $17.369$ \\
Planet Wind Density &$\rho_p$  & $5.881 \times 10^{-11} \mbox{g cm}^{-3}$ \\
Planet Mass Loss Rate &$\dot{M_p}$ & $10^{12} \mbox{g s}^{-1}$ \\
Stellar Mass &$M_\star$ & $1.35 M_{\astrosun}$ \footref{exoplanet} \\
Stellar Radius &$R_\star$ & $1.6 R_{\astrosun}$ \footref{exoplanet} \\
Stellar Wind Temperature &$T_\star$ & $10^6 \mbox{K}$ \\
Stellar Escape Parameter &$\lambda_\star$ & $19.542$ \\
Stellar Wind Density &$\rho_\star$  & $5.881 \times 10^{-15} \mbox{g cm}^{-3}$ \\
Stellar Mass Loss Rate &$\dot{M_{\star}}$ & $10^{-16} M_{\astrosun}/\mbox{yr}$ ($6\times10^9 g/s$) \\
Mass Ratio (planet/star) & $q$ & $9.92 \times 10^{-4}$ \\  
Orbital Separation & $a$ & $0.02293$ AU \footref{exoplanet} \\
Orbital Period & $P$ & $1.091$ days \footref{exoplanet} \\
 \hline
\end{tabular}
\end{minipage}
\end{table}

\subsection{Isothermal winds}

Both stellar and planetary winds are launched from fixed density and temperature boundaries at the surface of the star and planet. Because we set $\gamma = 1.0001$, we expect that they will expand isothermally as a Parker wind \citep{parker58}. The escape parameter $\lambda = \frac{G M \mu}{R k_B T}$ (see Table \ref{tab:parameters}) is a measure of the strength of this wind: for $\lambda \sim 10$, a Parker-type wind is expected, while for $\lambda \gg 10$, the planetary material is bound too tightly to escape. Since we are interested in the global distribution of the escaped planetary material, which is not expected to be significantly affected by the distribution and physical characteristics of the gas very close to the planet, the details of the planetary atmospheric escape can be disregarded. Simulations of the circumplanetary material would require a more detailed modelling of the physical processes driving the atmospheric escape and hence the modelling of the upper planetary atmosphere, possibly taking into account magnetic fields and non-thermal escape processes \citep[e.g.][]{ildar2016}.

We therefore initialize the stellar and planetary winds with the transonic Parker solution,
\begin{equation}
    \psi - \log \psi = 4 \log \xi + \frac{4}{\xi} - 3
  \label{eq:parker}
\end{equation}
with an analytic solution given by the Lambert W function
\begin{equation}
    \psi = 
      \begin{cases} 
       -W_0(\frac{-e^{3-\frac{4}{\xi}}}{\xi^4}) & \xi \leq 1 \\
       -W_{-1}(\frac{-e^{3-\frac{4}{\xi}}}{\xi^4}) & \xi > 1
      \end{cases}
\end{equation}
where $\psi = (\frac{v}{c_s})^2$ and $\xi = \frac{r}{r_s}$, with $r_s$ being the location where the transonic wind reaches the speed of sound.

More generally, the wind may be entirely subsonic (or supersonic), with
\begin{equation}
    \psi - \log \psi = 4 \log \xi + \frac{4}{\xi} + C
  \label{eq:parker_general}
\end{equation}
where the choice of $C$ determines the class of solution. For $C < -3$, the solutions are multivalued and thus unphysical; for $C > -3$, the principal branch $W_0$ gives subsonic solutions for all $\xi$, in which density and pressure build up at infinity, while the lower branch $W_{-1}$ gives supersonic solutions for all values of $\xi$. Since supersonic winds are not believed to occur at the stellar surface, these solutions can be ignored.

The density of the Parker wind is given by
\begin{equation}
    \phi = e^{\frac{-2}{\xi}(\xi-1)+\frac{1}{2}(1+\psi)}
  \label{eq:parker_density}
\end{equation}
where $\phi = \frac{\rho}{\rho_s}$, $\rho_s$ being the density at $\xi = 1$. With this density, the ram pressure of the wind is $P_{ram} = \phi \psi c_s^2 \rho_s$ and the thermal pressure is the usual $P_{therm} = \frac{\phi \rho_s}{m_H} k_b T_p$.

For our system parameters, the planet is located at $\xi_\star = 0.316$, well inside the sonic radius of the stellar wind. This agrees qualitatively with the conclusion of \citet{lai2010}, who find that for a solar-analog wind WASP-12b would be at $\xi = 0.719$.

\subsection{Description of simulation}

The model consists of a planet and star, modeled after the WASP-12 system, in a corotating reference frame. We assume that the planet is tidally locked, so that the orbital rotation rate, the planetary rotation rate, and the frame rotation rate are identical. For convenience, the stellar rotation rate is set equal to the orbital rotation rate, so that the stellar boundary remains fixed in the corotating reference frame. Because we assume the stellar wind conserves angular momentum, this means that the  velocity of the stellar wind in the orbital direction is approximately 8/9 that of the planet's velocity, giving them a relative velocity of $2.2\times10^6 < c_{s,\star} = 9\times10^6$. The corotating assumption is therefore in contrast with the fact that the orbital velocity of the planet is supersonic relative to the stellar wind \citep{vidotto2010}. However, in order for the stellar wind to suppress a significant portion of the planetary wind the bow shock (where the ram pressure of the planetary and stellar winds are estimated to be equal) must be within the Hill radius of the planet, so that the planetary material is unable to escape. Solving for the Parker wind density that gives $R_{Hill} \geq R_{bow}$, we find
\begin{equation}
    \rho_\star \geq 4.95\times10^{-11} \mbox{g cm}^{-3}
\end{equation}
which results in a mass loss rate for the star of
\begin{equation}
    \dot{M_\star} \geq 1.54\times10^{-12} M_{\astrosun}/\mbox{yr } (\sim1\times10^{14} g/s)
\end{equation}
which is 100x greater than the solar mass loss rate.

The locations of the star and planet are given by $\frac{-q}{q+1}a$ and $\frac{1}{q+1}a$, where $q=\frac{M_p}{M_\star}$. The simulation domain ranges from $[-3,-3,-1]\frac{a}{q+1}$ to $[3,3,1]\frac{a}{q+1}$, with a base resolution of $384 \times 384 \times 128$ and 2 levels of additional refinement, giving an effective resolution of $1536 \times 1536 \times 512$. The planetary radius is therefore resolved by nine cells.

The planetary wind has a density at the base of $\rho_p = 5.881 \times 10^{-11} \mbox{g cm}^{-3}$. We adopted this value to match the mass loss rate of $10^{12} \mbox{g s}^{-1}$ \citep{ehrenreich2011} predicted on the basis of the energy-limited approximation \citep{erkaev2007}. The stellar wind density $\rho_\star = 5.881 \times 10^{-15} \mbox{g cm}^{-3}$ produces a mass loss rate of $10^{-16} M_{\astrosun}/\mbox{yr}$. To simulate the effects of differential heating, we gave the planet an anisotropic temperature profile, $T(\theta) = T_p \max \left [ 0.01, \cos (\theta) \right ]$, where $\theta = 0$ and $\theta = \frac{\pi}{2}$ are the substellar and antistellar points, respectively \citep{carroll16}.

Both the star and planet are initialized with isothermal Parker winds, modified with contributions from the effects of rotation such that the velocities are given by
\begin{equation}
\boldsymbol{v}\left ( \boldsymbol{r} \right ) = 
\begin{cases}
        0 &  r \leq R_p \\
        v_p(r) \hat{\boldsymbol{r}} - \boldsymbol{\Omega} \times {\boldsymbol{r}} \left ( R_{p}^2/r^2 - 1\right) & r > R_p
\end{cases}
\label{eq:velocity_corrections}
\end{equation}
where $\Omega$ is again the orbital rotation rate. The stellar solution is initialized over the whole grid and held fixed out to $1.1 R_\star$, as well as in the simulation boundaries. The stellar wind is allowed to relax for 1 orbit, after which the planet and its initial wind solution are inserted out to the predicted bow shock radius $R_{bow} = 0.13a$. The planet's surface is held fixed at $T(\theta)$, the maximum value of which is assumed from calculations of the equilibrium temperature of a radiatively-launched planetary wind \citep{murrayclay09}, and $\rho_p$, and hence at a particular pressure, while the Parker solution is allowed to evolve freely for 14 more orbits.

\subsection{Magnesium column density} \label{sec:mg_column}

Only hydrogen is tracked in this simulation; however, we are primarily interested in ionized magnesium, since the presence of a disk was suggested by COS observations covering the core of the MgII\,h\&k resonance lines. We use the estimate of the MgII column density in the torus provided by \citet{haswell12} as a reference for the analysis of the simulation results. In order to estimate the MgII column density in our simulation, we consider the average hydrogen density $\rho$ taken over a cylinder stretching from the stellar boundary to the edge of the simulation, located at a distance of $0.062 \mbox{AU}$. We further consider a magnesium-to-hydrogen abundance ratio of $10^{-4}$ and assume that 10\% of the magnesium is singly ionized. This is a very conservative estimate, as magnesium is quickly ionized by the far-UV and EUV stellar radiation. This gives a column density of $\frac{\rho}{m_H} (10^{-4}) (10^{-1}) (2.68 a)\ \mbox{cm}^{-2}$. To arrive at the $2 \times 10^{17} \mbox{cm}^{-2}$ required to reproduce the observed absorption \citep{haswell12}, we therefore need the gas density in the torus produced by the planetary wind to reach an average value of $\rho = 3.65 \times 10^{-14} \mbox{g cm}^{-3}$.

\subsection{Radiation pressure} \label{sec:rad_poressure}

Our simulation does not include the effects of radiation pressure. We can, however, estimate the stellar Ly-$\alpha$ flux at which radiation pressure would become a significant contributor to the orbital dynamics of the torus. To estimate the radiation force per unit mass from Ly-$\alpha$, we can use $F_{rad} \sim \frac{f \pi e^2}{m_e c^2}\frac{h}{m_p}N_\alpha \sim 1.6 \times 10^{-20} N_\alpha$, where $ f=0.416$ is the oscillator strength for Ly-$\alpha$, $N_\alpha$ is the number of Ly-$\alpha$ photons per unit time per unit area, $m_e$ and $m_p$ are electron and proton masses, $h$ is Planck's constant, $c$ is the speed of light, and $e$ is the electron charge. 
 
For radiation pressure to push material completely out of orbit, it must exert a force per unit mass close to that of gravity. Taking the extreme case, where only radiation pressure is required to dissipate the disk, the condition is $F_{rad} > g = \frac{G M_\star}{a^2}$. For our system, $g = 1524 \mbox{cm}^2\mbox{ s}^{-1}$ at the orbital radius $a$. Using $F_{rad}$ above, the condition becomes $N_\alpha > \frac{1524}{1.6 \times 10^{-20}} = 9 \times 10^{22} \mbox{cm}^{-2} \mbox{ s}^{-1}$. We can compare this to the value for the sun, where $N_\alpha(1 \mbox{AU}) = 2.3 \times 10^{11} \mbox{cm}^{-2}\mbox{ s}^{-1}$, giving $N_\alpha(0.02293 \mbox{AU}) \sim 1.1 \times 10^{15}$. Thus, using solar values for the flux, we estimate that radiation pressure from Ly-$\alpha$ would not contribute significantly to the system dynamics.

\section{Results} \label{sec:result}

\begin{figure}
\centering
\includegraphics[width=\columnwidth]{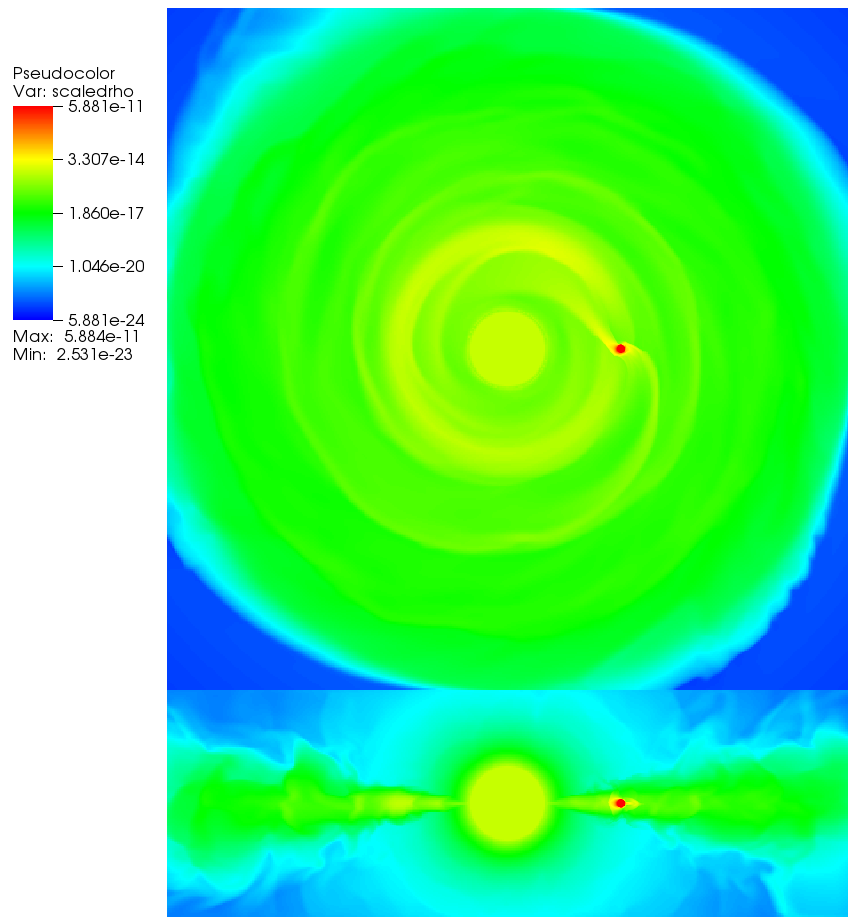}
\caption{Hydrogen density, in g\,cm$^{-3}$, at the final state of the simulation, shown with a horizontal (top) and vertical (bottom) cut of the simulation box through the line connecting the star and planet.}
\label{fig:simulation_end}
\end{figure}

\begin{figure}
\centering
\includegraphics[width=\columnwidth]{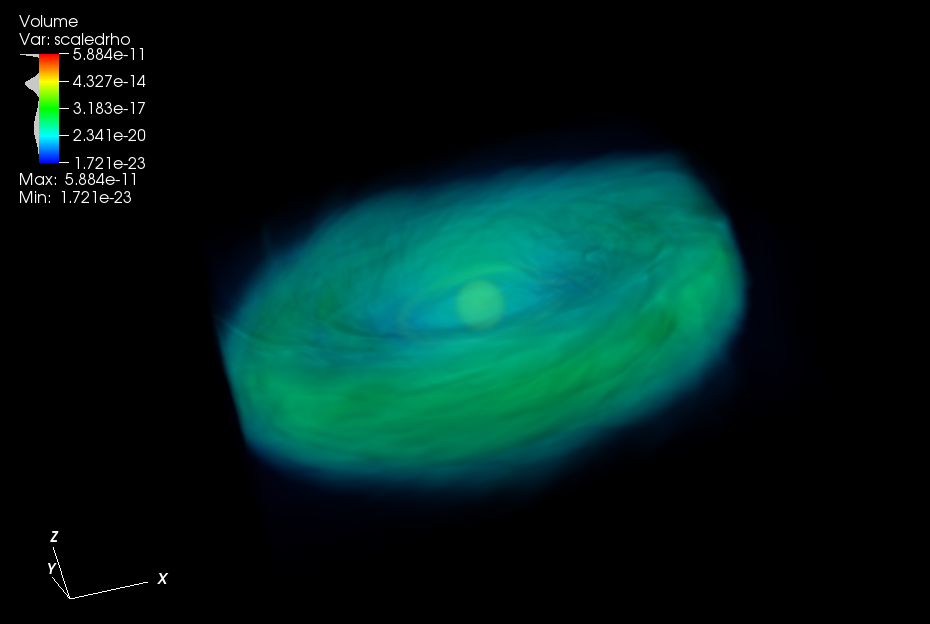}
\caption{Three-dimensional display of the final state of the simulation, showing the extent of the dense region of the disk surrounding WASP-12. The color code gives the hydrogen density in g\,cm$^{-3}$. Note how the disk covers the whole stellar surface as seen from a distant observer lying on the planetary orbital plane.}
\label{fig:volume_render}
\end{figure}

The final state of the system is shown in Figure \ref{fig:simulation_end}, with a 3D visualization showing the final disk and star in Figure \ref{fig:volume_render}. The planet is located along the x axis in the three-dimensional plot, but cannot be seen due to its small size. 

We find that a dense gaseous torus does form around WASP-12. Figure \ref{fig:simulation_end} shows that the densest gas (yellow color scales) forms a ring ($\rho = 10^{-14} \mbox{g cm}^{-3}$) that extends over almost the entire circumference of the planet's orbit. Extending both inward and outward from the ring we also see high density gas ($\rho = 10^{-17} \mbox{g cm}^{-3}$) forming an extended disk or torus that reaches all the way to the edge of the simulation box. We note that the outer (radial) edges of the torus as seen in the simulations are smoothed by the outflow-only boundary conditions. The torus is relatively thin in z and $T_{disk} = T_p$ due to the low value of the polytropic index $\gamma$, where $P = \rho^\gamma$.

The stellar wind is driving into the planetary wind during the entire simulation, but is unable to sweep away the gas in the torus. This occurs because of the small value of $\dot{M_\star}$ chosen; in addition, because the planet orbits within the stellar wind's sonic radius the pressure of the stellar wind is dominated by thermal pressure, meaning it is at least two orders of magnitude less than the pressure of the planetary wind.

At the end of the simulation, which covers 14 orbits, the disk has an average density over the cylinder described in Section \ref{sec:mg_column} of $1.29 \times 10^{-16} \mbox{g cm}^{-3}$, giving a MgII column density of $7.09\times10^{14} \mbox{cm}^{-2}$.  This is not yet high enough to yield the MgII absorption observed in \citet{haswell12} and \citet{fossati13}, which required a MgII column density of $2 \times 10^{17} \mbox{cm}^{-2}$. However, as we show below, the mass in the torus continues to build with time.

\subsection{Evolution over time}

\begin{figure}
\centering
\includegraphics[width=\columnwidth]{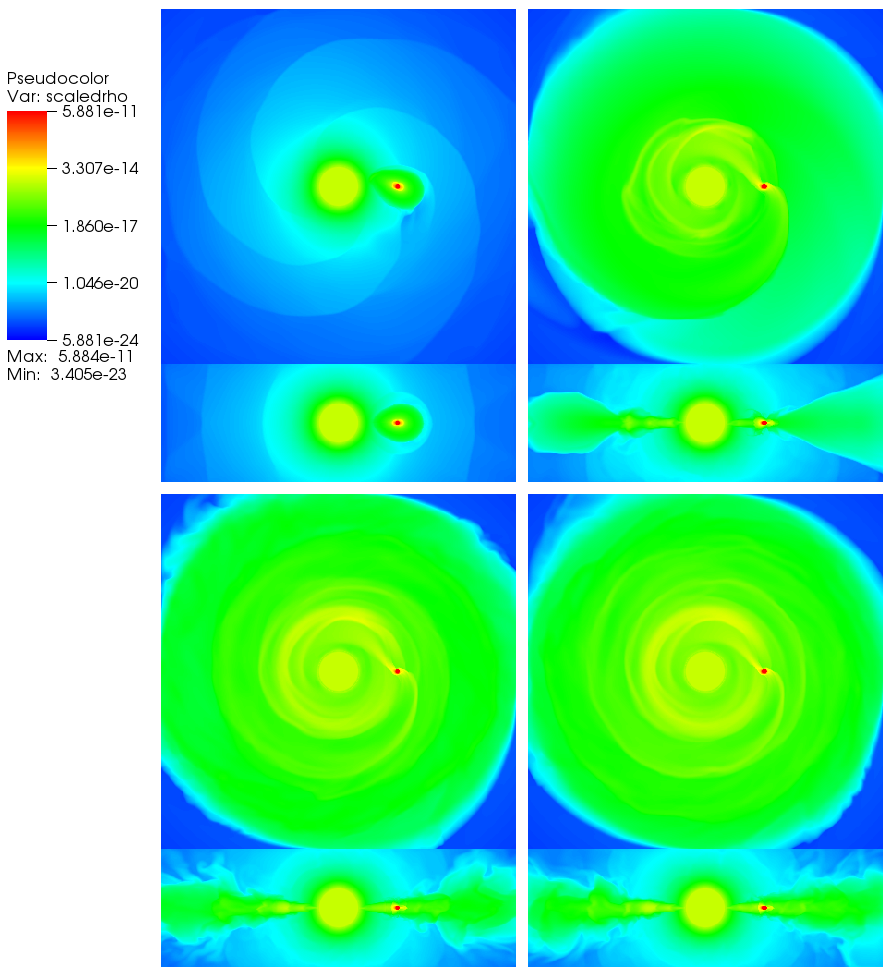}
\caption{Same as Fig. \ref{fig:simulation_end}, but after zero planetary orbits (top-left), 2 planetary orbits (top-right), 9 planetary orbits (bottom-left), and 14 planetary orbits (bottom-right). A gas disk starts forming immediately after the first revolution, and is firmly established after the second orbit.}
\label{fig:simulation_over_time}
\end{figure}

\begin{figure}
\centering
\includegraphics[width=\columnwidth]{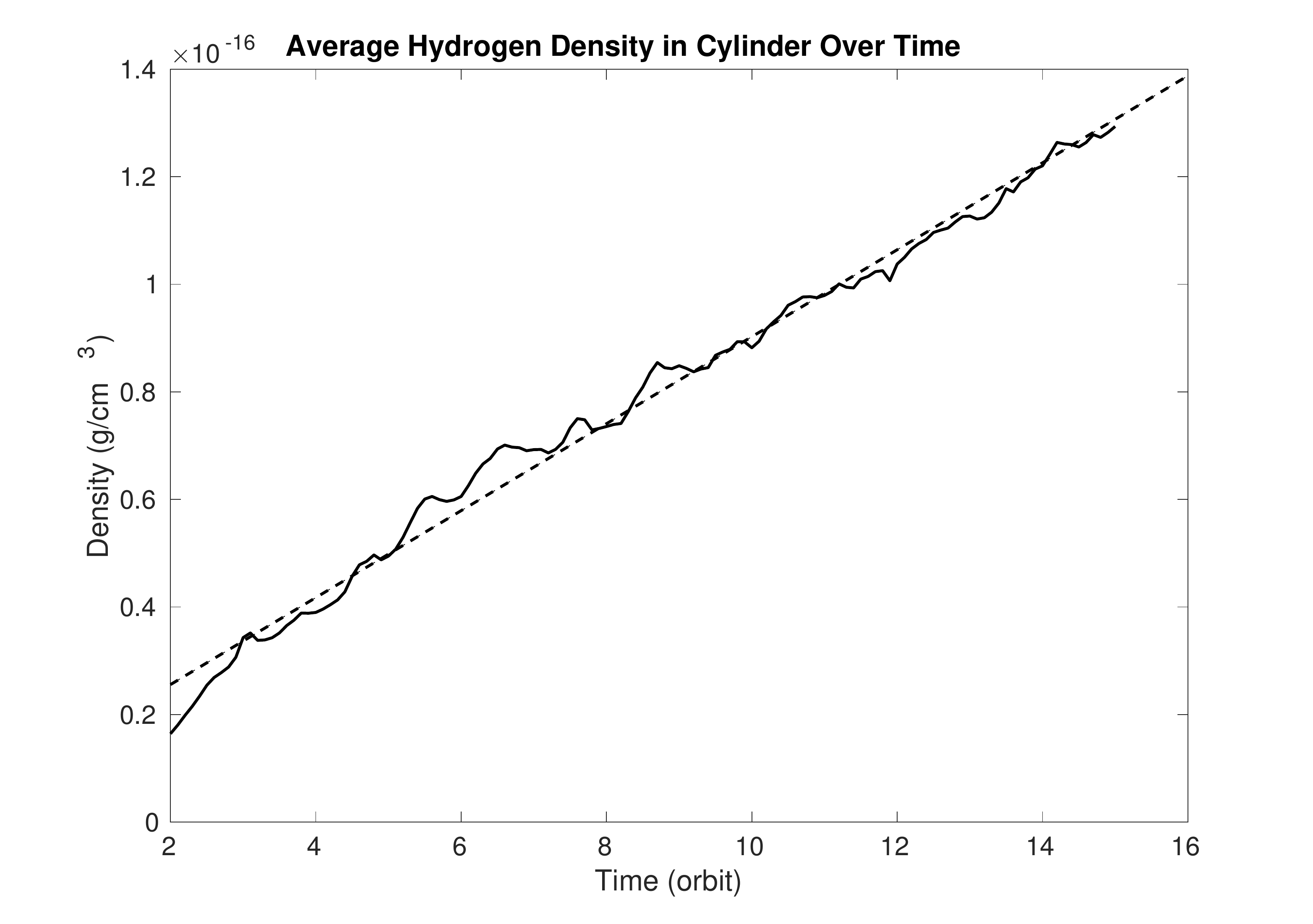}
\caption{Average density of hydrogen within the cylinder described in Sect. \ref{sec:mg_column}, with a linear fit shown by the dashed line. Although the required density of $10^{-14} \mbox{g cm}^-3$ is not reached within the time-frame of the simulation, extrapolating the linear fit to longer times leads to the required MgII column density in about 13 years. While we predict that the disk will eventually reach a steady state with a roughly constant density, the linear fit is a good approximation for short timescales.}
\label{fig:cyl_density_over_time}
\end{figure}

In terms of the flow morphology, the system reaches a quasi-steady state after the planet's second orbit (Fig. \ref{fig:simulation_over_time}). Importantly, though, we see that the disk density is rising throughout the simulation. We observe an approximately linear rate of increase of $\dot{\rho}$ = $8.09 \times 10^{-18} \mbox{g cm}^{-3}$orbit$^{-1}$, as shown in Fig. \ref{fig:cyl_density_over_time}. We note that the growth rate does appear to slow slightly, likely due to a slight decrease in the planetary mass loss rate. This is probably due to the increasing pressure of the disk, which alters the downstream boundary condition for the planet's Parker-type wind.

We can use this linear approximation for $\dot{\rho}$ to project the increasing mass of the disk forward. If the observed trend continued, it would take approximately 13 years for the density of the disk to be sufficient to create the observed MgII absorption. With such a short disk formation timescale, even periods of high stellar activity would not remove the disk for a long time. If the disk were disrupted by a large increase in the stellar wind, we would expect that the disk would reform quickly once the stellar mass loss rates returned to lower values.

The torus will eventually come into a steady state configuration where the mass input from the planetary wind will be balanced by mass loss through its boundaries (both vertical and radial). In order to approximate this final state of the planetary wind, we chose the constant $C$ in the spherical Parker solution (Eq. \ref{eq:parker_general}) such that the total pressure of the wind ($P_w = P_{ram} + P_{therm}$) is equal to the pressure of the disk at the desired density, $P_d = n_H k_b T_p$, with $n_H$ being the number density of hydrogen required to reach a MgII column density of $2 \times 10^{17} \mbox{cm}^{-2}$, at a distance equal to the height of the disk from the surface of the planet. This occurs when $C = 21.9151$, resulting in a maximum wind speed (at $\xi = 1$) of $2.15 \mbox{cm s}^{-1}$. In fact, the disk pressure alone does not become sufficient to shut off the wind completely until a density 2000x higher than that required for the observed absorption.

Although the wind pressure at the surface of the planet is always greater than $P_d$, we assume that a combination of mass input from the wind and mass loss from the outer edges of the torus will eventually cause the disk to reach a quasi-steady state. Therefore, the torus will not build up enough mass to become optically thick at all wavelengths, which would contradict the observations. However, it would be necessary to run the simulation for approximately 10 million CPU hours (about 100000 years wall-time, using the same cluster) to test this assumption.

\subsection{Synthetic Observations}

\begin{figure*}
\centering
\includegraphics[width=\textwidth]{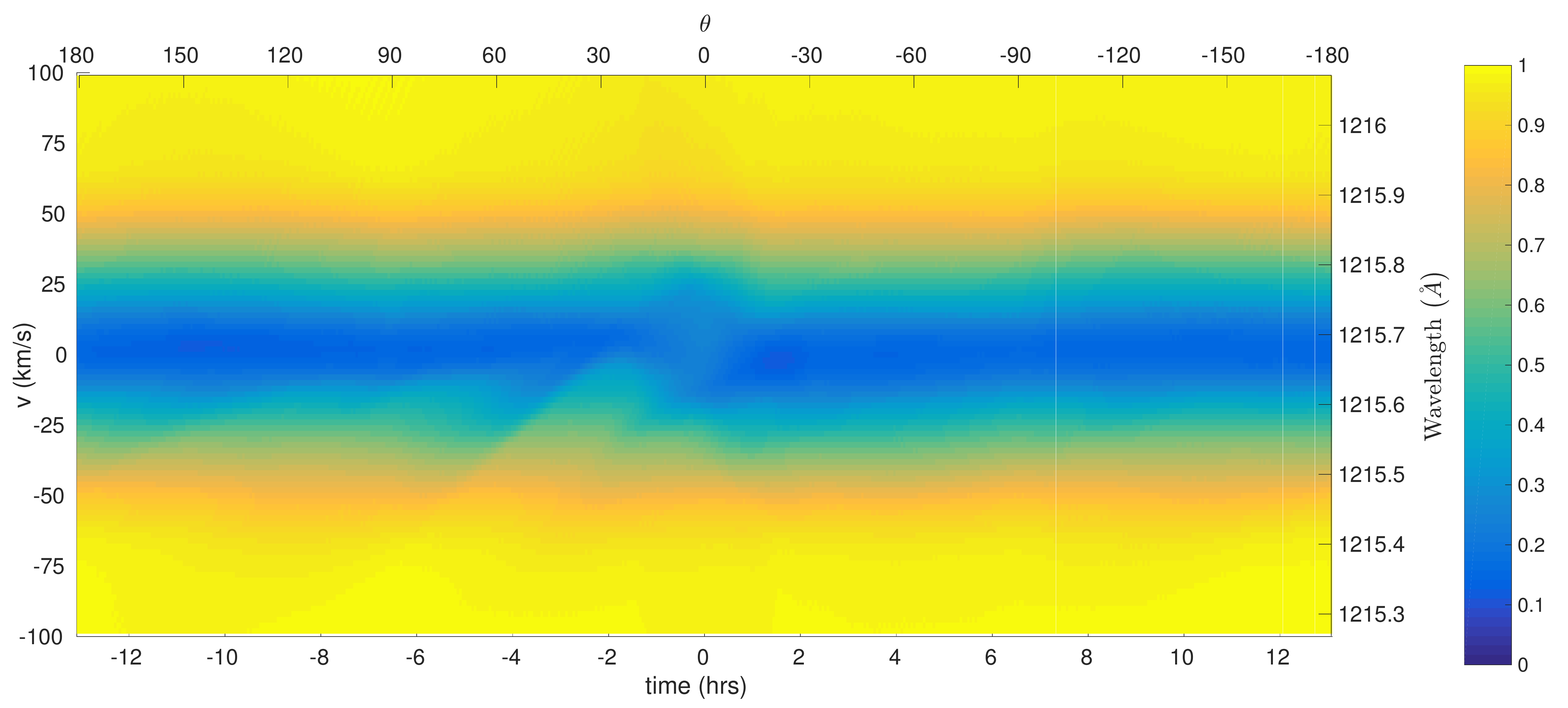}
\caption{Average normalized Ly alpha intensity over the last 5 simulated orbits as seen from a distant observer lying on the planet's orbital plane. The intensity is shown as a function of wavelength (right axis) or velocity (left axis) and also as a function of time from mid-transit (bottom axis) or orbital phase angle \citep[top axis; see Fig. 11 of][for more details]{carroll16}. The Ly-$\alpha$ line is strongly absorbed out to $\pm50 \mbox{km s}^{-1}$, regardless of the planet's orbital phase.}
\label{fig:attenuation}
\end{figure*}

To quantify the amount of stellar light absorbed by the disk, we calculate the optical depth as a function of frequency as 

\begin{equation}
    \tau (\nu) = \sigma_{\nu_0} \int{ds \int{d\nu' n (\nu') \phi (\nu - \nu')}}
\end{equation}

where the normalization constant of the absorption cross section profile for neutral hydrogen $\sigma_{\nu_0} = 1.102 \times 10^{-2} \mbox{cm}^2 \mbox{s}^{-1}$ \citep{bourrier13} and the line integration is along rays emanating from the surface of the star towards an observer.  Since each cell has a particular velocity, the distribution of density per frequency for cell $j$ is 

\begin{equation}
 n(\nu') = n_j \delta(\nu'-\nu_j) \mbox{ where } \nu_j = \nu_0(1 - \frac{\mathbf{v}_j \cdot \hat{n}}{c}),
\end{equation}

where $n_j$ is number density of cell $j$. This gives us

\begin{equation}
    \tau(\nu) = \sigma_{\nu_0} \int{ds \, n_j \phi(\nu - \nu_j)}\,.
\end{equation}
Then, by integrating over frequency bin $i$, we get
\begin{equation}
    \tau_i \Delta \nu = \int_{\nu_i-\frac{\Delta \nu}{2}}^{\nu_i+\frac{\Delta \nu}{2}} \tau(\nu) d\nu = \sigma_{\nu_0} \int{ds \, n_j \int_{\nu_i-\frac{\Delta \nu}{2}}^{\nu_i+\frac{\Delta \nu}{2}} \phi(\nu - \nu_j)}\,.
\end{equation}

We further assume that the line profile is from thermal broadening and approximate it as 

\begin{equation}
    \phi(\nu - \nu_j) = \frac{1}{\sqrt{\pi} \Delta \nu_D} \exp^{-(\nu-\nu_j)^2/(\Delta \nu_D)^2}\,,
\end{equation}
where $\Delta \nu_D = \frac{\nu_0}{c} \sqrt{\frac{2 k_B T}{m}}$.

The integral then becomes
\begin{equation}
    \hspace{-0.5cm}
    \tau_i = \frac{\sigma_{\nu_0}}{2\Delta \nu} \int{ds \, n_j \left[ \erf{\left( \frac{\nu_j - \nu_i + \frac{1}{2} \Delta \nu}{\Delta \nu_D} \right)} - \erf {\left( \frac{\nu_j-\nu_i-\frac{1}{2} \Delta \nu}{\Delta \nu_D} \right)} \right]}\,.
\end{equation}

This can also be defined by an effective absorption coefficient per cell $j$ 
per frequency bin $i$ of
\begin{equation}
    \alpha_{j,i}  =  \frac{n_j\sigma_{\nu_0}}{2 \Delta \nu} \left[ \erf{\left( \frac{\nu_j - \nu_i + \frac{1}{2} \Delta \nu}{\Delta \nu_D} \right)} - \erf {\left( \frac{\nu_j-\nu_i-\frac{1}{2} \Delta \nu}{\Delta \nu_D} \right) }\right]\,,
\end{equation}
which can be integrated along rays emanating from the surface of the star using standard ray tracing techniques
\begin{equation}
  \tau_i = \int{\alpha_{j,i}\, ds}.
\end{equation}

The simulation data was saved every 1/10th of an orbit. For each frame, sets of rays emanating from the surface of the star were integrated through the simulation domain at different angles to create synthetic images with a spatial resolution of $200^2$ at each frequency and angle of observation. The frequency resolution corresponds to $2 \mbox{km s}^{-1}$ and the angle resolution corresponds to $3 \mbox{ min}$ of orbital time. These images were then summed to calculate a normalized intensity as a function of frequency, angle, and simulation time.  To the extent that the fluid flow is steady, different viewing angles for a fixed simulation time can be translated to different observational times. Finally, for each frequency and viewing angle, we calculated the mean of the normalized intensity over the last 5 orbits, as seen in Fig. \ref{fig:attenuation}. 

Because of the large interstellar medium absorption which occurs over the large distance to the star \citep[388$\pm$37\,pc;][]{gaia}, and of the torus absorption, the Ly-$\alpha$ line of WASP-12 cannot be directly observed, even with future facilities (e.g. LUVOIR). The same is also true for the stellar X-ray and far-UV emission, which would have otherwise helped to interpret the origin of the anomalous, apparent lack of stellar activity \citep[e.g.][]{fossati2015b}. As a result, although radial velocity measurements suggest that WASP-12 has an activity typical of other stars of similar temperature \citep{bonomo2017}, the actual intensity of the stellar chromospheric emission is unknown; therefore, our synthetic observations cannot be compared to actual observations. However, key aspects of our results can be helpful in interpreting the lack of MgII\,h\&k line core emission.

Figure \ref{fig:attenuation} shows that after just a few revolutions of the planet, the gas in the disk significantly absorbs the entire central part of the Ly-$\alpha$ line, regardless of the planet's orbital phase. The complete absorption on the line center is in agreement with the COS observations, which showed that the emission cores of the MgII\,h\&k resonance lines were always completely absent, even a few hours before the transit \citep{haswell12,nichols2015}. Note that the emission/absorption map shown in Fig. \ref{fig:attenuation} is that obtained towards the end of our simulation run, while the density of the gas in the disk is still increasing. This implies that the absorption would increase considerably if the simulation had been allowed to run for many years, as required to reach the observed MgII column density.

\section{Discussion and conclusion} \label{sec:discuss}

The WASP-12 system hosts one of the most inflated and irradiated known hot Jupiters. Ultraviolet observations of WASP-12 indicate the presence of an anomalous lack of the ubiquitous emission at the core of the MgII\,h\&k resonance lines.  In this paper we examine the hypothesis that the emission lines are absorbed via a translucent circumstellar disk or torus formed by material lost by during photoevaporation of the planet's atmosphere \citep{haswell12,fossati13}. To determine if the hot Jupiter WASP-12b is responsible for generating a gaseous disk dense enough to absorb the expected stellar emission at the core of the MgII\,h\&k lines, we simulate the complete WASP-12 system for 15 orbits.

Our simulations followed the fluid-dynamical evolution of a wind driven off of an evaporating planet.  Because of the short period, Coriolis forces drive the wind into "up-orbit" and "down-orbit" arms, which leads to the development of a circumstellar disk or torus over time \citep{carroll16}. Tracing the disk over time showed a continuous, nearly linear rate of increase in disk density. By extrapolating the simulations forward in time, we estimate that it will take approximately 13 years ($\approx$4350 planetary orbits) for the disk to reach a high enough density to absorb stellar MgII\,h\&k lines, thereby reproducing the observations. We note that this time could be somewhat larger because of the decrease in the planetary mass loss rate caused by the external pressure of the disk. Even if this increase required a few orders of magnitude longer duration, however, it would still be insignificant compared to the age of the system.

Two factors not included in this study may have an impact on the disk structure and density. Magnetic fields have been shown to reduce the mass loss rates from hot Jupiters by an order of magnitude or more \citep{owen14,khodachenko15}, leading to roughly the same increase in the time required to build up a dense enough disk. As noted above, this would still be significantly less than the age of the system. 

Of greater consequence is radiation pressure on the disk material. On the one hand, radiation pressure from the star could drive the gas escaping the planet out of the planetary orbital plane, thus reducing its density. On the other hand, if the radiation pressure from the star were to balance the disk pressure at a distance larger than the planet's orbital separation, the planet could have an increased mass loss rate because of the loss of the external pressure from the disk A higher mass loss could leading in the end to an increase in the disk density. These effects, particularly that of radiation pressure, should be examined in further studies.

In conclusion, we find the hypothesis that a torus formed from a photoevaporating planet wind can account for the lack of MgII lines plausible. Future work should explore both the fluid dynamics and radiative transfer of the process in more detail.

\section{Acknowledgments}

This work used the computational and visualization resources in the Center for Integrated Research Computing (CIRC) at the University of Rochester. Financial support for this project was provided by the Department of Energy grant GR523126, the National Science Foundation grant GR506177, and the Space Telescope Science Institute grant GR528562.

\bibliography{WASP12.bib}

\begin{thebibliography}{}
\makeatletter
\relax
\def\mn@urlcharsother{\let\do\@makeother \do\$\do\&\do\#\do\^\do\_\do\%\do\~}
\def\mn@doi{\begingroup\mn@urlcharsother \@ifnextchar [ {\mn@doi@}
  {\mn@doi@[]}}
\def\mn@doi@[#1]#2{\def\@tempa{#1}\ifx\@tempa\@empty \href
  {http://dx.doi.org/#2} {doi:#2}\else \href {http://dx.doi.org/#2} {#1}\fi
  \endgroup}
\def\mn@eprint#1#2{\mn@eprint@#1:#2::\@nil}
\def\mn@eprint@arXiv#1{\href {http://arxiv.org/abs/#1} {{\tt arXiv:#1}}}
\def\mn@eprint@dblp#1{\href {http://dblp.uni-trier.de/rec/bibtex/#1.xml}
  {dblp:#1}}
\def\mn@eprint@#1:#2:#3:#4\@nil{\def\@tempa {#1}\def\@tempb {#2}\def\@tempc
  {#3}\ifx \@tempc \@empty \let \@tempc \@tempb \let \@tempb \@tempa \fi \ifx
  \@tempb \@empty \def\@tempb {arXiv}\fi \@ifundefined
  {mn@eprint@\@tempb}{\@tempb:\@tempc}{\expandafter \expandafter \csname
  mn@eprint@\@tempb\endcsname \expandafter{\@tempc}}}

\bibitem[\protect\citeauthoryear{{Bisikalo}, {Kaygorodov}, {Ionov},
  {Shematovich}, {Lammer}  \& {Fossati}}{{Bisikalo} et~al.}{2013}]{bisikalo13}
{Bisikalo} D.,  {Kaygorodov} P.,  {Ionov} D.,  {Shematovich} V.,  {Lammer} H.,
   {Fossati} L.,  2013, \mn@doi [\apj] {10.1088/0004-637X/764/1/19}, 764, 19

\bibitem[\protect\citeauthoryear{{Bonomo}, {Desidera}, {Benatti}  \& et
  al.}{{Bonomo} et~al.}{2017}]{bonomo2017}
{Bonomo} A.~S.,  {Desidera} S.,  {Benatti} S.,   et al. 2017, preprint, \href
  {http://adsabs.harvard.edu/abs/2017arXiv170400373B} {} (\mn@eprint {arXiv}
  {1704.00373})

\bibitem[\protect\citeauthoryear{{Bourrier} \& {Lecavelier des
  Etangs}}{{Bourrier} \& {Lecavelier des Etangs}}{2013}]{bourrier13}
{Bourrier} V.,  {Lecavelier des Etangs} A.,  2013, \mn@doi [\aap]
  {10.1051/0004-6361/201321551}, 557, A124

\bibitem[\protect\citeauthoryear{Carroll-Nellenback, Shroyer, Frank  \&
  Ding}{Carroll-Nellenback et~al.}{2013}]{carroll13}
Carroll-Nellenback J.~J.,  Shroyer B.,  Frank A.,   Ding C.,  2013, \mn@doi
  [Journal of Computational Physics]
  {http://dx.doi.org/10.1016/j.jcp.2012.10.004}, 236, 461

\bibitem[\protect\citeauthoryear{{Carroll-Nellenback}, {Frank}, {Liu},
  {Quillen}, {Blackman}  \& {Dobbs-Dixon}}{{Carroll-Nellenback}
  et~al.}{2017}]{carroll16}
{Carroll-Nellenback} J.,  {Frank} A.,  {Liu} B.,  {Quillen} A.~C.,  {Blackman}
  E.~G.,   {Dobbs-Dixon} I.,  2017, \mn@doi [MNRAS] {10.1093/mnras/stw3307},
  466, 2458

\bibitem[\protect\citeauthoryear{{Cunningham}, {Frank}, {Varni{\`e}re},
  {Mitran}  \& {Jones}}{{Cunningham} et~al.}{2009}]{cunningham09}
{Cunningham} A.~J.,  {Frank} A.,  {Varni{\`e}re} P.,  {Mitran} S.,   {Jones}
  T.~W.,  2009, \mn@doi [\apjs] {10.1088/0067-0049/182/2/519}, \href
  {http://adsabs.harvard.edu/abs/2009ApJS..182..519C} {182, 519}

\bibitem[\protect\citeauthoryear{{Ehrenreich} \& {D{\'e}sert}}{{Ehrenreich} \&
  {D{\'e}sert}}{2011}]{ehrenreich2011}
{Ehrenreich} D.,  {D{\'e}sert} J.~M.,  2011, \mn@doi [\aap]
  {10.1051/0004-6361/201016356}, \href
  {http://adsabs.harvard.edu/cgi-bin/bib_query?arXiv:1103.0011} {529, A136}

\bibitem[\protect\citeauthoryear{{Ehrenreich} et~al.,}{{Ehrenreich}
  et~al.}{2012}]{ehrenreich2012}
{Ehrenreich} D.,  et~al., 2012, \mn@doi [\aap] {10.1051/0004-6361/201219981},
  \href {http://adsabs.harvard.edu/abs/2012A%26A...547A..18E} {547, A18}

\bibitem[\protect\citeauthoryear{{Erkaev}, {Kulikov}, {Lammer}, {Selsis},
  {Langmayr}, {Jaritz}  \& {Biernat}}{{Erkaev} et~al.}{2007}]{erkaev2007}
{Erkaev} N.~V.,  {Kulikov} Y.~N.,  {Lammer} H.,  {Selsis} F.,  {Langmayr} D.,
  {Jaritz} G.~F.,   {Biernat} H.~K.,  2007, \mn@doi [\aap]
  {10.1051/0004-6361:20066929}, \href
  {http://adsabs.harvard.edu/abs/2007A%26A...472..329E} {472, 329}

\bibitem[\protect\citeauthoryear{{Fossati} et~al.,}{{Fossati}
  et~al.}{2010a}]{fossati10a}
{Fossati} L.,  et~al., 2010a, \mn@doi [\apjl] {10.1088/2041-8205/714/2/L222},
  714, L222

\bibitem[\protect\citeauthoryear{{Fossati} et~al.,}{{Fossati}
  et~al.}{2010b}]{fossati2010b}
{Fossati} L.,  et~al., 2010b, \mn@doi [\apj] {10.1088/0004-637X/720/1/872},
  \href {http://adsabs.harvard.edu/abs/2010ApJ...720..872F} {720, 872}

\bibitem[\protect\citeauthoryear{{Fossati}, {Ayres}, {Haswell}, {Bohlender},
  {Kochukhov}  \& {Fl{\"o}er}}{{Fossati} et~al.}{2013}]{fossati13}
{Fossati} L.,  {Ayres} T.~R.,  {Haswell} C.~A.,  {Bohlender} D.,  {Kochukhov}
  O.,   {Fl{\"o}er} L.,  2013, \mn@doi [\apjl] {10.1088/2041-8205/766/2/L20},
  766, L20

\bibitem[\protect\citeauthoryear{{Fossati}, {Ingrassia}  \& {Lanza}}{{Fossati}
  et~al.}{2015a}]{fossati2015a}
{Fossati} L.,  {Ingrassia} S.,   {Lanza} A.~F.,  2015a, \mn@doi [\apjl]
  {10.1088/2041-8205/812/2/L35}, \href
  {http://adsabs.harvard.edu/abs/2015ApJ...812L..35F} {812, L35}

\bibitem[\protect\citeauthoryear{{Fossati}, {France}, {Koskinen}, {Juvan},
  {Haswell}  \& {Lendl}}{{Fossati} et~al.}{2015b}]{fossati2015b}
{Fossati} L.,  {France} K.,  {Koskinen} T.,  {Juvan} I.~G.,  {Haswell} C.~A.,
  {Lendl} M.,  2015b, \mn@doi [\apj] {10.1088/0004-637X/815/2/118}, \href
  {http://adsabs.harvard.edu/abs/2015ApJ...815..118F} {815, 118}

\bibitem[\protect\citeauthoryear{{Gaia Collaboration} et~al.,}{{Gaia
  Collaboration} et~al.}{2016}]{gaia}
{Gaia Collaboration} et~al., 2016, \mn@doi [\aap]
  {10.1051/0004-6361/201629272}, \href
  {http://cdsads.u-strasbg.fr/abs/2016A%26A...595A...1G} {595, A1}

\bibitem[\protect\citeauthoryear{{Haswell} et~al.,}{{Haswell}
  et~al.}{2012}]{haswell12}
{Haswell} C.~A.,  et~al., 2012, \mn@doi [\apj] {10.1088/0004-637X/760/1/79},
  \href {http://adsabs.harvard.edu/abs/2012ApJ...760...79H} {760, 79}

\bibitem[\protect\citeauthoryear{{Hebb} et~al.,}{{Hebb}
  et~al.}{2009}]{hebb2009}
{Hebb} L.,  et~al., 2009, \mn@doi [\apj] {10.1088/0004-637X/693/2/1920}, \href
  {http://adsabs.harvard.edu/abs/2009ApJ...693.1920H} {693, 1920}

\bibitem[\protect\citeauthoryear{{Khodachenko}, {Shaikhislamov}, {Lammer}  \&
  {Prokopov}}{{Khodachenko} et~al.}{2015}]{khodachenko15}
{Khodachenko} M.~L.,  {Shaikhislamov} I.~F.,  {Lammer} H.,   {Prokopov} P.~A.,
  2015, \mn@doi [\apj] {10.1088/0004-637X/813/1/50}, 813, 50

\bibitem[\protect\citeauthoryear{{Koskinen}, {Harris}, {Yelle}  \&
  {Lavvas}}{{Koskinen} et~al.}{2013a}]{koskinen2013a}
{Koskinen} T.~T.,  {Harris} M.~J.,  {Yelle} R.~V.,   {Lavvas} P.,  2013a,
  \mn@doi [\icarus] {10.1016/j.icarus.2012.09.027}, \href
  {http://adsabs.harvard.edu/abs/2013Icar..226.1678K} {226, 1678}

\bibitem[\protect\citeauthoryear{{Koskinen}, {Yelle}, {Harris}  \&
  {Lavvas}}{{Koskinen} et~al.}{2013b}]{koskinen2013b}
{Koskinen} T.~T.,  {Yelle} R.~V.,  {Harris} M.~J.,   {Lavvas} P.,  2013b,
  \mn@doi [\icarus] {10.1016/j.icarus.2012.09.026}, \href
  {http://adsabs.harvard.edu/abs/2013Icar..226.1695K} {226, 1695}

\bibitem[\protect\citeauthoryear{{Koskinen}, {Lavvas}, {Harris}  \&
  {Yelle}}{{Koskinen} et~al.}{2014}]{koskinen2014}
{Koskinen} T.~T.,  {Lavvas} P.,  {Harris} M.~J.,   {Yelle} R.~V.,  2014,
  \mn@doi [Philosophical Transactions of the Royal Society of London Series A]
  {10.1098/rsta.2013.0089}, \href
  {http://adsabs.harvard.edu/abs/2014RSPTA.37230089K} {372, 20130089}

\bibitem[\protect\citeauthoryear{{Lai}, {Helling}  \& {van den Heuvel}}{{Lai}
  et~al.}{2010}]{lai2010}
{Lai} D.,  {Helling} C.,   {van den Heuvel} E.~P.~J.,  2010, \mn@doi [\apj]
  {10.1088/0004-637X/721/2/923}, 721, 923

\bibitem[\protect\citeauthoryear{{Lanza}}{{Lanza}}{2014}]{lanza2014}
{Lanza} A.~F.,  2014, \mn@doi [\aap] {10.1051/0004-6361/201425051}, \href
  {http://adsabs.harvard.edu/abs/2014A%26A...572L...6L} {572, L6}

\bibitem[\protect\citeauthoryear{{Lecavelier des Etangs} et~al.,}{{Lecavelier
  des Etangs} et~al.}{2010}]{lecavelier10}
{Lecavelier des Etangs} A.,  et~al., 2010, \mn@doi [\aap]
  {10.1051/0004-6361/200913347}, 514, A72

\bibitem[\protect\citeauthoryear{{Lecavelier des Etangs} et~al.,}{{Lecavelier
  des Etangs} et~al.}{2012}]{lecavelier12}
{Lecavelier des Etangs} A.,  et~al., 2012, \mn@doi [\aap]
  {10.1051/0004-6361/201219363}, \href
  {http://cdsads.u-strasbg.fr/abs/2012A\%26A...543L...4L} {543, L4}

\bibitem[\protect\citeauthoryear{{Linsky}, {Yang}, {France}, {Froning},
  {Green}, {Stocke}  \& {Osterman}}{{Linsky} et~al.}{2010}]{linsky10}
{Linsky} J.~L.,  {Yang} H.,  {France} K.,  {Froning} C.~S.,  {Green} J.~C.,
  {Stocke} J.~T.,   {Osterman} S.~N.,  2010, \mn@doi [\apj]
  {10.1088/0004-637X/717/2/1291}, \href
  {http://adsabs.harvard.edu/abs/2010ApJ...717.1291L} {717, 1291}

\bibitem[\protect\citeauthoryear{{Matsakos}, {Uribe}  \&
  {K{\"o}nigl}}{{Matsakos} et~al.}{2015}]{matsakos15}
{Matsakos} T.,  {Uribe} A.,   {K{\"o}nigl} A.,  2015, \mn@doi [\aap]
  {10.1051/0004-6361/201425593}, 578, A6

\bibitem[\protect\citeauthoryear{{Murray-Clay}, {Chiang}  \&
  {Murray}}{{Murray-Clay} et~al.}{2009}]{murrayclay09}
{Murray-Clay} R.~A.,  {Chiang} E.~I.,   {Murray} N.,  2009, \mn@doi [\apj]
  {10.1088/0004-637X/693/1/23}, \href
  {http://adsabs.harvard.edu/abs/2009ApJ...693...23M} {693, 23}

\bibitem[\protect\citeauthoryear{{Nichols} et~al.,}{{Nichols}
  et~al.}{2015}]{nichols2015}
{Nichols} J.~D.,  et~al., 2015, \mn@doi [\apj] {10.1088/0004-637X/803/1/9},
  \href {http://adsabs.harvard.edu/abs/2015ApJ...803....9N} {803, 9}

\bibitem[\protect\citeauthoryear{{Owen} \& {Adams}}{{Owen} \&
  {Adams}}{2014}]{owen14}
{Owen} J.~E.,  {Adams} F.~C.,  2014, \mn@doi [MNRAS] {10.1093/mnras/stu1684},
  444, 3761

\bibitem[\protect\citeauthoryear{{Parker}}{{Parker}}{1958}]{parker58}
{Parker} E.~N.,  1958, \mn@doi [\apj] {10.1086/146579}, \href
  {http://adsabs.harvard.edu/abs/1958ApJ...128..664P} {128, 664}

\bibitem[\protect\citeauthoryear{{Schneiter}, {Vel{\'a}zquez}, {Esquivel},
  {Raga}  \& {Blanco-Cano}}{{Schneiter} et~al.}{2007}]{Schneiter2007}
{Schneiter} E.~M.,  {Vel{\'a}zquez} P.~F.,  {Esquivel} A.,  {Raga} A.~C.,
  {Blanco-Cano} X.,  2007, \mn@doi [\apjl] {10.1086/524945}, \href
  {http://adsabs.harvard.edu/abs/2007ApJ...671L..57S} {671, L57}

\bibitem[\protect\citeauthoryear{{Schneiter}, {Esquivel}, {D'Angelo},
  {Vel{\'a}zquez}, {Raga}  \& {Costa}}{{Schneiter}
  et~al.}{2016}]{Schneiter2017}
{Schneiter} E.~M.,  {Esquivel} A.,  {D'Angelo} C.~S.~V.,  {Vel{\'a}zquez}
  P.~F.,  {Raga} A.~C.,   {Costa} A.,  2016, \mn@doi [\mnras]
  {10.1093/mnras/stw076}, \href
  {http://adsabs.harvard.edu/abs/2016MNRAS.457.1666S} {457, 1666}

\bibitem[\protect\citeauthoryear{{Shaikhislamov} et~al.,}{{Shaikhislamov}
  et~al.}{2016}]{ildar2016}
{Shaikhislamov} I.~F.,  et~al., 2016, \mn@doi [\apj]
  {10.3847/0004-637X/832/2/173}, \href
  {http://adsabs.harvard.edu/abs/2016ApJ...832..173S} {832, 173}

\bibitem[\protect\citeauthoryear{{Vidal-Madjar}, {Lecavelier des Etangs},
  {D{\'e}sert}, {Ballester}, {Ferlet}, {H{\'e}brard}  \&
  {Mayor}}{{Vidal-Madjar} et~al.}{2003}]{vidalmadjar03}
{Vidal-Madjar} A.,  {Lecavelier des Etangs} A.,  {D{\'e}sert} J.-M.,
  {Ballester} G.~E.,  {Ferlet} R.,  {H{\'e}brard} G.,   {Mayor} M.,  2003,
  \mn@doi [\nat] {10.1038/nature01448}, \href
  {http://adsabs.harvard.edu/abs/2003Natur.422..143V} {422, 143}

\bibitem[\protect\citeauthoryear{{Vidal-Madjar} et~al.,}{{Vidal-Madjar}
  et~al.}{2004}]{vidalmadjar04}
{Vidal-Madjar} A.,  et~al., 2004, \mn@doi [\apjl] {10.1086/383347}, \href
  {http://adsabs.harvard.edu/abs/2004ApJ...604L..69V} {604, L69}

\bibitem[\protect\citeauthoryear{{Vidal-Madjar} et~al.,}{{Vidal-Madjar}
  et~al.}{2013}]{vidalmadjar13}
{Vidal-Madjar} A.,  et~al., 2013, \mn@doi [\aap] {10.1051/0004-6361/201322234},
  560, A54

\bibitem[\protect\citeauthoryear{{Vidotto}, {Jardine}  \& {Helling}}{{Vidotto}
  et~al.}{2010}]{vidotto2010}
{Vidotto} A.~A.,  {Jardine} M.,   {Helling} C.,  2010, \mn@doi [\apjl]
  {10.1088/2041-8205/722/2/L168}, \href
  {http://adsabs.harvard.edu/abs/2010ApJ...722L.168V} {722, L168}

\bibitem[\protect\citeauthoryear{{Vidotto}, {Bisikalo}, {Fossati}  \&
  {Llama}}{{Vidotto} et~al.}{2015}]{vidotto2015}
{Vidotto} A.~A.,  {Bisikalo} D.~V.,  {Fossati} L.,   {Llama} J.,  2015, in
  {Lammer} H.,  {Khodachenko} M.,  eds,  Astrophysics and Space Science Library
  Vol. 411, Characterizing Stellar and Exoplanetary Environments. p.~153,
  \mn@doi{10.1007/978-3-319-09749-7_8}

\makeatother
\end{thebibliography}

\bsp

\label{lastpage}

\end{document}